\documentclass[ams,prl,twocolumn,amssymb,superscriptaddress,longbibliography,floatfix]{revtex4-2}
\usepackage{amsmath}
\usepackage{tabularx}
\usepackage{bm}
\usepackage{slashed}
\usepackage{euscript}
\usepackage{graphicx}
\usepackage{xcolor,colortbl}
\usepackage{amsfonts}
\usepackage{exscale}
\usepackage{colortbl}
\usepackage{amsbsy}
\usepackage{subfigure}
\usepackage{textcomp}
\usepackage{comment}
\usepackage{mathrsfs}
\usepackage{afterpage}

\usepackage[normalem]{ulem} %underlines OFF in hyperref

\usepackage{bbold}
\usepackage{blkarray}

\usepackage{ulem}
\usepackage[final]{pdfpages}
\usepackage{pgffor}

\makeatletter
\AtBeginDocument{\let\LS@rot\@undefined}
\makeatother

\usepackage[colorlinks,citecolor=blue,linkcolor=red,urlcolor=blue]
{hyperref}

\newcommand{\be}{\begin{equation}}
\newcommand{\ee}{\end{equation}}

\begin{document}

% \title{Dispersion and damping of the amplitude Schmid-Higgs mode \\ in disordered superconductors \\
% TENTATIVE: Schmid-Higgs (amplitude) mode in disordered superconductors} 
\title{Spatially-resolved dynamics of the amplitude Schmid-Higgs mode \\
in disordered superconductors } 

\author{P. A. Nosov}
  % \altaffiliation{These authors contributed equally to this work.}
\affiliation{Department of Physics, Harvard University, Cambridge, Massachusetts 02138, USA}

\author{E. S. Andriyakhina}
\affiliation{Freie Universit\"at Berlin, Dahlem Center for Complex Quantum Systems and Fachbereich Physik, Arnimallee 14, Berlin, 14195, Germany}

\author{I. S. Burmistrov}
 \affiliation{L.D. Landau Institute for Theoretical Physics, acad. Semenova av.1-a, Chernogolovka,
142432, Russia}
\affiliation{Laboratory for Condensed Matter Physics, HSE University, Moscow, 101000, Russia}

	\date{\today}
	
    \begin{abstract}
    We investigate the spatially-resolved dynamics of the collective amplitude Schmid-Higgs (SH) mode in disordered $s$-wave superconductors and fermionic superfluids. By analyzing the analytic structure of the zero-temperature SH susceptibility in the complex frequency plane, we find that when the coherence length greatly exceeds the mean free path: (i) the SH response at fixed wave vectors exhibits late-time oscillations decaying as $1/t^2$ with frequency $2\Delta$, where $\Delta$ is the superconducting gap; (ii) sub-diffusive oscillations with a dynamical exponent $z{=}4$ emerge at late times and large distances; and (iii) spatial oscillations at fixed frequency decay exponentially, with a period that diverges as the frequency approaches $2\Delta$ from above. When the coherence length is comparable to the mean free path, additional exponentially-decaying oscillations at fixed wave vectors appear with frequency above $2\Delta$. Furthermore, we show that the SH mode induces an extra peak in the third-harmonic generation current at finite wave-vectors. The frequency of this peak is shifted from the conventional resonance at $\Delta$, thereby providing an unambiguous signature of order parameter amplitude dynamics.
	\end{abstract}

	\maketitle
%\textbf{Introduction.\,--} 
\noindent{Investigating} the collective excitations in superconductors provides crucial insights into the complex structure of their order parameter and associated dynamical responses \cite{vaks1962collective,schmid1968,Artemenko1979,Kulik1981,Arseev2006,Shimano2019}. Unlike the well-studied phase fluctuations, the collective dynamics of the order parameter amplitude (so-called Schmid-Higgs (SH) mode) has received much less attention due to experimental challenges in its detection, primarily caused by its decoupling from density fluctuations. However, recent advances in terahertz and Raman spectroscopic probes have made direct observation of the SH mode more accessible \cite{SHTerahertz2013,SH_measured_2014,sherman2015higgs,Grasset2018,Kota2024}, in turn prompting a renewed wave of theoretical interest in the amplitude fluctuations \cite{Kos2004,Combescot2006,Barankov2006,Yuzbashyan2006,Dzero_2009,Podolsky2011,Gazit2013,Naoto2015,Cea2016,moor2017amplitude,Fischer2018,Shen2018,Kurkjian2019,Murotani2019,Silaev2019,Sun2020,Lee2023,Phan2023,PhysRevLett.134.066002}.

The properties of the SH mode in a disorder-free limit are relatively well established in both three-dimensional (3D) \cite{AndrianovPopov,Carlson1973,Podolsky2011,Kurkjian2019} and two-dimensional (2D) \cite{Podolsky2011,Gazit2013,Phan2023} systems, across weak and strong coupling regimes. However, real materials inevitably contain impurities or other structural imperfections, making it imperative to understand how disorder influences the fluctuations of the superconducting order parameter. Despite extensive research of collective responses in dirty superconductors \cite{Smith1995,Reizer2000,Mondal2011,Cea2014,Cea2015,Shtyk2017,Silaev2019,kamenev2023field,Dzero2024,Dzero_2_2024,2024BoFan,wang2024higgs}, a comprehensive description of the spatially-resolved SH dynamics in this limit is still lacking. In particular, the dispersion relation and the associated long-distance and late-time oscillatory behavior of the SH mode in the presence of disorder remain unknown. Another open question is how the SH mode contributes to the nonlinear current response at finite wave vectors.

The goal of the present Letter is to fill this gap by examining the non-analyticities of the $T{=}0$ disorder-averaged SH susceptibility $\chi_{\textsf{SH}}(z,\bm{q})$ as a function of complex frequency and momentum. This susceptibility quantifies the dynamical response of the order parameter amplitude $|\Delta(t,\bm{r})|$ that arises when external perturbations disturb it from its equilibrium state. Our approach is based on
the BCS model with impurities, assuming a local attractive coupling $\lambda$ that induces an $s$-wave, spin-singlet superconducting state with a zero-temperature gap $\Delta$. For concreteness, we consider the following Hamiltonian density
\begin{equation}\label{eq:H}
    \mathcal{H}{=} \sum\limits_{\sigma}\psi^\dagger_{\sigma}\Big(-\frac{(\nabla{-}i\bm{A})^2}{2m}{-}\mu {+}V(\bm{r})\Big)\psi_{\sigma} -\frac{\lambda}{\nu}\psi^\dagger_{\uparrow}\psi^\dagger_{\downarrow}\psi_{\downarrow}\psi_{\uparrow}
\end{equation}
Here $\sigma{=}{\uparrow}{/}{\downarrow}$ labels spin degrees of freedom, $\nu$ is the density of states at the Fermi level in the normal state, $\lambda{>}0$ is the dimensionless BCS coupling constant, $\psi_\sigma(\bm{r})$ is the electron field, $\bm{A}$ is the vector potential, and $V(\bm{r})$ is a white-noise Gaussian random potential that induces elastic scattering rate $1/\tau$. We assume that both $1/\tau$ and the {superconducting} gap $\Delta$ are much smaller than the Fermi energy, {$\mu{=}mv_F^2/2$ ($v_F$ is the Fermi velocity and $m$ is the electron mass)}, which allows us to linearize the electron dispersion, and treat impurity scattering within the self-consistent Born approximation. The ratio $\Delta \tau$ is used to interpolate between the dirty ($\Delta\tau{\ll} 1$) and clean ($\Delta \tau {\gg} 1$) regimes. Under these conditions, the interplay of superconductivity and disorder is treated at the level of Anderson's theorem \cite{Anderson1959,Gor'kovAbrikosov1959a,Gor'kovAbrikosov1959b}, disregarding more subtle effects such as interference-induced corrections \cite{Maekawa1981,Maekawa1984,Finkelstein1987,Finkelstein1994}, or spatial inhomogeneity of the order parameter and Lifshitz tails below the spectral edge \cite{larkin1972density,Meyer2001,Feigelman2012}.

\noindent\textsf{\textcolor{blue}{SH susceptibility.}} The collective dynamics of the superconducting order parameter is described by the standard Ginzburg-Landau functional obtained from Eq.~\eqref{eq:H} at $\bm{A}{=}0$, with its quadratic part encoding Gaussian fluctuations of the amplitude $|\Delta|$ around its mean-field value~\footnote{The coupling between the amplitude and phase fluctuations of the order parameter $\Delta=|\Delta|e^{i\theta}$ vanishes after averaging over disorder in the BCS limit due to the effective particle-hole symmetry~\cite{Smith1995,andriyakhina2024quantum}.}
\begin{equation}\label{eq:def_chi}
    \begin{aligned}
    &S_{\Delta\Delta}=\int_{\bm{r}}\sum\limits_n\Delta(i\omega_n,\bm{r})\chi^{-1}_{\textsf{SH}}(i\omega_n,\bm{r},\bm{r}')\Delta(-i\omega_n,\bm{r}') \\
    &\chi^{-1}_{\textsf{SH}}(i\omega_n,\bm{r},\bm{r}') = \lambda^{-1}\delta(\bm{r}-\bm{r}')-\Pi_{\Delta\Delta}(i\omega_n,\bm{r},\bm{r}'),
    \end{aligned}
\end{equation}
where $\omega_n{=}2\pi Tn$ is the bosonic Matsubara frequency, $\chi_{\textsf{SH}}$ is the real space Matsubara SH susceptibility in a given disorder realization, and $\Pi_{\Delta\Delta}(i\omega_n,\bm{r},\bm{r}')$ is the Fourier transform of the imaginary time correlation function {$(\pi\nu)^{-1}\langle \mathcal{T} \hat\Delta(\tau,\bm{r})\hat\Delta(0,\bm{r}') \rangle$}. 
Here $\hat\Delta(\tau,\bm{r})$ is the $s$-wave, spin-singlet order parameter amplitude.
   
The expectation value is taken in the standard BCS state with the uniform mean-field order parameter $\Delta$ determined via the gap equation $1{=}\pi \lambda T \sum_m 1/E_{\varepsilon_m}$. Here $E_{\varepsilon_m}{=}\sqrt{\varepsilon_m^2{+}\Delta^2}$, and $\varepsilon_m{=}2\pi T(m{+}1/2)$ is the fermionic Matsubara frequency.

In this setup, the disorder-averaged Matsubara SH susceptibility $\chi_{\textsf{SH}}(i\omega_n,\bm{q})$ is given by \cite{SM}
\begin{equation}
\label{eq:SH_Matsubara}
    \frac{1}{\chi_{\textsf{SH}}}{=}  \pi T \sum_{m}\biggl\{ \frac{1}{E_{\varepsilon_m}}{-}S_{\bm{q}}\left(E_{\varepsilon_m}{+}E_{\tilde{\varepsilon}_m}\right)\\
    \biggl [1{+}\frac{\varepsilon_m \tilde{\varepsilon}_m{-}\Delta^2}{E_{\varepsilon_m}E_{\tilde{\varepsilon}_m}}
    \biggr] \biggr \} ,
\end{equation}
where $\tilde{\varepsilon}_m{=}\varepsilon_m{+}|\omega_n|$. This expression assumes %momenta 
{$q{\ll}mv_F$,} but fully captures the crossover between the diffusive and ballistic scales through the structure factor $S_{\bm{q}}\left(E\right)$. It is given by $S^{\textsf{2D}}_{\bm{q}}\left(E\right){=} 1/(|\mathcal{E}|{-}1/\tau)$ for a 2D system and by $S^{\textsf{3D}}_{\bm{q}}\left(E \right){=}1/(v_F q/\arg \mathcal{E}{-}1/\tau)$ in 3D, where $\mathcal{E}{=}E{+}1/\tau{+}iv_F q$. In the diffusive limit, $v_Fq$, $E {\ll} 1/\tau$, 
the structure factor in Eq.~\eqref{eq:SH_Matsubara} reduces to \cite{Kulik1981,Smith1995,andriyakhina2024quantum} 
\begin{equation} \label{eq:S_d}
    S_{\bm{q}}(E) = 1/(Dq^2+E), \quad D{=}v_F^2\tau/d,
\end{equation}
where $d{=}2,3$ is the dimensionality. At $q{=}0$, Eq.~\eqref{eq:SH_Matsubara} reduces to its clean limit for any $\Delta\tau$ as a manifestation of Anderson's theorem.

\color{black}

%The calculated behavior in Eqs.~\eqref{eq:omega-r-FT}-\eqref{eq:z=4_oscillations}) agrees with the numerical results in Fig.~\ref{fig:fig2}(a,b). 

%The spatially-resolved properties of the SH susceptibility 

% In particular, both real and imaginary parts of $\chi^R$ can be extracted from Eq.~\eqref{eq:current}. 

% Below, we show how these results are derived from the analytic structure of $\chi_{\textsf{SH}}$ in the complex frequency plane.

% In both (ii) and (iii), the oscillations envelope decay exponentially, with a power-law prefactor that depends on the dimensionality. 

\color{black}

%%%%%%%%%%%%%%%%%%%%%%%%%%%%%%%%%%%
\begin{figure}[t!]
\center{\includegraphics[width=0.95\linewidth]{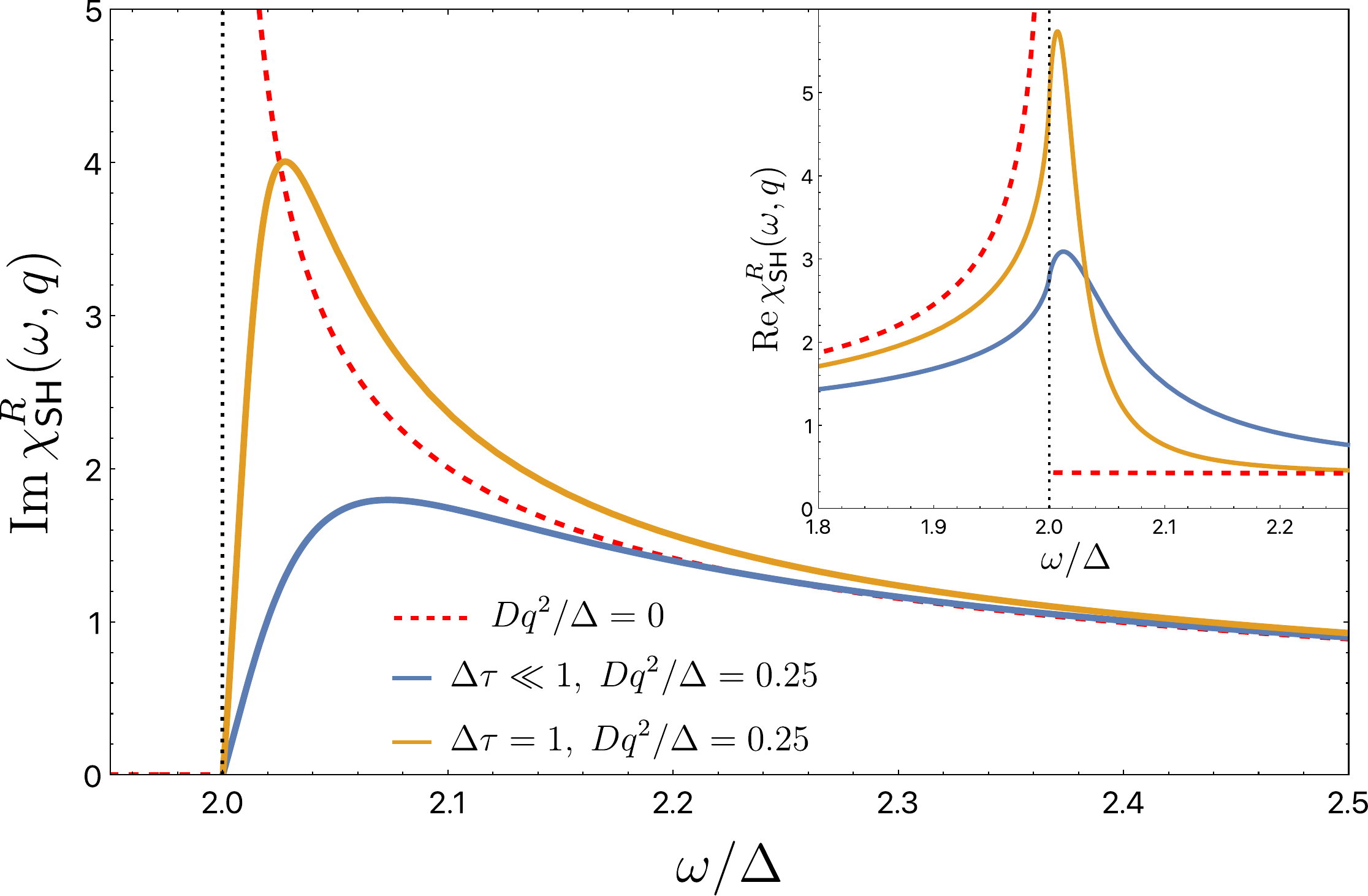}}
\caption{The spectral function $\operatorname{Im}\chi_{\textsf{SH}}^{R}(\omega,\bm{q})$ for $Dq^2/\Delta{=}0.25$ ($\xi q{=}0.5$). The blue (orange) curve corresponds to $\Delta \tau{\ll} 1$ ($\Delta\tau{=}1$) in 2D. The dotted line indicates the continuum edge. The dashed red line corresponds to $q{=}0$ independent of $\Delta \tau$. Inset: $\operatorname{Re}\chi_{\textsf{SH}}^{R}(\omega,\bm{q})$ for the same parameters.} 
\label{fig:fig1}
\end{figure}
%%%%%%%%%%%%%%%%%%%%%%%%%%%%%%%%%%%

%%%%%%%%%%%%%%%%%%%%%%%%%%%%%%%%%%%
\begin{figure*}[t!]
\center{\includegraphics[width=1.0\textwidth]{Fig2_V2.png}}
\caption{(a) $\operatorname{Im}\chi_{\textsf{SH}}^{\downarrow}(z,\bm{q})$ (orange surface) in the complex frequency plane for $Dq^2/\Delta{=}0.5$ and $\Delta\tau{\ll} 1$. The red solid line corresponds to the real axis $\operatorname{Im}z{=}0$, and the red vertical region indicates the discontinuity in $\operatorname{Im}\chi_{\textsf{SH}}^{\downarrow}(z,\bm{q})$ along its branch cut. The transparent blue plane marks zero on the vertical axis. (b) The  contour plot of $\operatorname{Im}\chi_{\textsf{SH}}^{\downarrow}(z,\bm{q})$ for the same parameters. The red dashed line indicated the branch cut of $\chi_{\textsf{SH}}^{\downarrow}(z,\bm{q})$. The yellow dashed line shows the trajectory of the pole while momentum is varied. 
(c) The frequency of the SH mode as a function of $\xi^2q^2{\equiv} Dq^2/\Delta$ for $\Delta \tau {\ll} 1$ (left panel) and $\Delta \tau {=} 1$ (right panel) in 2D, shown with a yellow solid curve. The small-$q$ asymptotic behavior, 
Eq.~\eqref{eq:z_q_final}, is shown with the dashed red curve. The normalized spectral function $2\arctan(\operatorname{Im}\chi_{\textsf{SH}}^R(\omega,\bm{q}))/\pi {\in} [0,1]$ is shown with the background color. The yellow dashed line shows the position of the maximum of the spectral function. The horizontal white dashed line denotes the edge of the two-particle continuum. The SH damping rate 
(blue solid curve) is shown at the bottom. }
\label{fig:fig2}
\end{figure*}
%%%%%%%%%%%%%%%%%%%%%%%%%%%%%%%%%%%

% \noindent\textsf{\textcolor{blue}{Analytic continuation and the SH mode.}} 

The retarded SH susceptibility $\chi_{\textsf{SH}}^R(\omega,\bm{q})$ at $T{=}0$ is obtained from Eq.~\eqref{eq:SH_Matsubara} in a standard way \cite{SM}. The branch cuts are chosen along the real axis such that for any real $|E|{\geq} \Delta$ we have $\sqrt{\Delta^2{-}(E{\pm}i 0)^2}{=}{\mp} i \operatorname{sgn}(E)\sqrt{E^2{-}\Delta^2}$. \color{black}If extended to the lower half-plane as well, the above choice of the branch cut defines the \emph{physical} Riemann sheet of $\chi_{\textsf{SH}}(z,\bm{q})$, with its imaginary part changing discontinuously across the cut along the real axis at $|\operatorname{Re}z|{\geq} 2\Delta$, and vanishing for $\operatorname{Im}z{=}0,\;|\operatorname{Re}z|{\leq} 2\Delta$ due to the presence of the superconducting gap (see Fig.~\ref{fig:fig1}). The resulting Cauchy representation of $[\chi_{\textsf{SH}}(z,\bm{q})]^{-1}$ on the physical sheet is given by
\begin{equation}\label{eq:SH_complex}
\frac{1}{\chi_{\textsf{SH}}}\hspace{-0.1em}=\hspace{-0.2em}\int\limits_{{-}\infty}^{+\infty}\hspace{-0.1em}\frac{d\varepsilon}{\pi} \hspace{-0.1em}\left[
    \frac{ \rho_{\bm{q}}(|\varepsilon|) \operatorname{sgn}\varepsilon}{\varepsilon{-}z}{+}\frac{\pi/2}{
   \sqrt{\varepsilon^2{-}4\Delta^2}
    }\right]\hspace{-0.2em}\theta\Bigl(\frac{|\varepsilon|}{2\Delta}{-}1\Bigr)
\end{equation}
for $z{\in} \mathbb{C}/\{{z}{:} \operatorname{Im}{z}{=}0, |\operatorname{Re}{z}|{\geq} 2\Delta\}$. The second term in Eq.~\eqref{eq:SH_complex} arises because $1/|\chi_{\textsf{SH}}(\omega_n,\bm{q})|$ is an increasing function of $\omega_n$, see \cite{SM} for details. This relation allows us to immediately interpret the spectral density $\operatorname{sgn}({\varepsilon})\theta(|{\varepsilon}|{-}2\Delta) \rho_{\bm{q}}(|{\varepsilon}|)$ in Eq.~\eqref{eq:SH_complex} as $\operatorname{Im}[1/\chi_{\textsf{SH}}^R({\varepsilon},\bm{q})]$, whereas the real part is obtained by taking the principal value of the integral. The full 
expression for $\rho_{\bm{q}}(\omega)$ is given in \cite{SM}, and its diffusive limit at $\Delta\tau{\ll} 1$ yields 
\begin{align}
   \rho_{\bm{q}}(\omega)&= \frac{4 \bar{\omega} ^2-\left(\bar{q}^4+\bar{\omega} ^2\right)^2}{\bar{q}^2 (\bar{\omega}{+}2 ) \left(\bar{q}^4{+}\bar{\omega} ^2\right)} \Pi \bigg(\hspace{-0.4em}\left.\frac{(\bar{\omega}{-}2 )^2 \left(\bar{q}^4{+}\bar{\omega} ^2\right)}{\bar{q}^4\left(4 -\bar{q}^4-\bar{\omega} ^2\right)}\right|\frac{\bar{\omega} -2  }{\bar{\omega}+2 }\bigg)\notag \\
     &+\frac{\bar{q}^2\left(4 +\bar{q}^4+\bar{\omega} ^2\right)}{  (\bar{\omega}+2 ) \left(\bar{q}^4+\bar{\omega} ^2\right)}  K\left(\frac{\bar{\omega} -2 }{\bar{\omega} +2 }\right).\label{eq:Pi_c_m_im}
     \end{align}
Here, we defined dimensionless variables $\bar{\omega}{=}\omega/\Delta$, $ \bar{q}{=}\xi q$, and $\xi{=}\sqrt{D/\Delta}$ is the coherence length. Also, $\Pi(x|y){=}\int_{0}^{\pi/2}d\alpha/ ((1{-}x\sin^2\alpha)\sqrt{1{-}y^2\sin^2\alpha})$ is the complete elliptic integral of the 3rd kind, and $K(x){=} \Pi(0|x)$. 
Eq.~\eqref{eq:Pi_c_m_im} only assumes 
$|\omega|, v_F q{\ll}1/\tau$, 
but $\xi q$ and $|\omega|/\Delta$ can be arbitrary.

The imaginary and real parts of $\chi_{\textsf{SH}}^R(\omega,\bm{q})$ at ${q}{>}0$ feature a peak at a frequency above $2\Delta$ for arbitrary values of $\Delta\tau$, as shown in Fig.~\ref{fig:fig1}. This peak shifts to higher frequencies when momentum is increased. At ${q}{=}0$, the peak is replaced with a square-root singularity at $\omega{=}2\Delta$ \cite{vaks1962collective}. Further assuming the dirty limit, $\xi q {\ll} 1$ and $0{\leq} \omega{-}2\Delta{\ll} \Delta$, while keeping the ratio $\Delta\xi^4q^4/(\omega{-}2\Delta) $ fixed (i.e. anticipating the $z{=}4$ dynamical exponent, defined by the relation $|\omega{-}2\Delta|{\sim}q^{z}$), we obtain 
\begin{equation}
\label{eq:chi_approx_cont}
\frac{1}{\chi_{\textsf{SH}}^R}{\simeq} \frac{\bar{q}^2}{4}
\Bigl [ \ln \frac{2^6}{\bar{q}^4}
{-} \!\sum_{s{=}\pm} (1{+}s u)\ln(u{+}s) {+} i \pi (1{-}u)
\Bigr ],
\end{equation}
where $u{=}\sqrt{1{+}4(\bar{\omega}{-}2)/\bar{q}^4}{\geq}1$, From Eq.~\eqref{eq:chi_approx_cont}, we find that the frequency and the width of the peak in $\chi^R_{\textsf{SH}}$ at $\xi q{\ll}1$ \cite{SM} scale as 
\begin{equation}
\label{eq:susceptibility_peak_width}
    \frac{\omega_{\rm max}(\bm{q})}{\Delta} {\approx}2{+} \frac{4\xi^4 q^4}{\pi^2} \ln^2 \frac{2\sqrt{2}}{\xi q}, \quad
    \frac{\gamma_{\rm max}(\bm{q})}{\Delta}{\sim} \xi^4 q^4 \ln \frac{1}{\xi q}.
\end{equation}
To calculate the late-time and long-distance behavior of the SH susceptibility, we now proceed to identify its nonanalyticities (e.g., poles) in the complex plane.

\noindent\textsf{\textcolor{blue}{SH mode as a pole in the SH susceptibility.}} The appearance of a peak in $\chi_{\textsf{SH}}^R(\omega,\bm{q})$ on the real frequency axis is already indicative of a pole in the lower half-plane. \color{black} However, as emphasized above, the presence of the branch cut implies that $\chi_{\textsf{SH}}(\omega{+}i0^+,\bm{q})$ is not smoothly connected to $\chi_{\textsf{SH}}(\omega{-}i0^+,\bm{q})$. In fact, $\chi_{\textsf{SH}}(z,\bm{q})$ does not have any non-analyticities in the lower half-plane on the physical Riemann sheet. Instead, one has to smoothly continue it through the branch cut into the \emph{unphysical} Riemann sheet, and search for a pole there \cite{AndrianovPopov,Kurkjian2019,Phan2023}. The resulting susceptibility, denoted as $\chi^{\downarrow}_{\textsf{SH}}(z,\bm{q})$, coincides with Eq.~\eqref{eq:SH_complex} in the upper half-plane, but remains continuous across the interval $\operatorname{Im}z{=}0,\; \operatorname{Re}z{>}2\Delta$. The structure of $\chi^\downarrow_{\textsf {SH}}(z,\bm{q})$ is demonstrated in Fig.~\ref{fig:fig2}(a,b), and the explicit formula for it is given in End Matter. Numerical evaluation reveals that $\chi^\downarrow_{\textsf {SH}}(z,\bm{q})$ indeed has a pole $z_{\bm{q}}$ in the lower half-plane at any finite momentum $q$.  The SH mode’s dispersion is then obtained as 
    $\omega_{\textsf{SH}}(\bm{q}){\equiv} \operatorname{Re}z_{\bm{q}}$ and $\gamma_{\textsf{SH}}(\bm{q}){\equiv}|\operatorname{Im} z_{\bm{q}}|$,
with both of these quantities exhibiting strong dependence on $\Delta \tau$
(see Fig.~\ref{fig:fig2}(c)). In the dirty limit $\Delta\tau{\ll} 1$, for $\xi q{\ll}1$ we  obtain
\begin{equation}
\label{eq:dispersion_damping}
    \frac{\omega_{\textsf{SH}}(\bm{q})}{\Delta} {\approx}2{-} \frac{4\xi^4 q^4}{\pi^2}\ln^2\frac{2\sqrt{\pi}} {\xi q},\quad 
    \frac{\gamma_{\textsf{SH}}(\bm{q})}{\Delta}{\approx}\frac{4 \xi^4 q^4}{\pi} \ln \frac{2\sqrt{\pi}} {\xi q}
\end{equation}
The $z{=}4$ scaling of $\omega_{\textsf{SH}}(\bm{q})$ with momentum can be also estimated directly from $S_{\bm{q}}(E)$ in Eq.~\eqref{eq:S_d} since the SH mode involves quasiparticles with energy $\omega {\gtrsim}\Delta$. Expanding $S_{\bm{q}}(\sqrt{\omega^2{-}\Delta^2})$ for such $\omega$, we find that the pole occurs at $|\omega{-}\Delta|{\sim}  D^2q^4/\Delta $ in the dirty limit. We also emphasize that $\omega_{\textsf{SH}}(\bm{q}){<}2\Delta$.  In the terminology of Ref.~\cite{Klein2020}, the pole is ``hidden" behind the branch cut of $\chi^\downarrow_{\textsf {SH}}(z,\bm{q})$ on the real axis at $\operatorname{Re}z{<}2\Delta$ (see Fig.~\ref{fig:fig2}(b,c)). Upon increasing $\Delta \tau$ while keeping $\xi q$ fixed, the pole $z_{\bm{q}}$ shifts to the right. Eventually, its frequency $\omega_{\textsf{SH}}(\bm{q})$ exceeds $2\Delta$ — that is, the pole becomes ``visible" — and it contributes to the Fourier transform, giving rise to additional exponentially decaying oscillations at frequency $\omega_{\textsf{SH}}(\bm{q})$ in $\chi^R_{\textsf{SH}}(t,\bm{q})$ (cf. Eq.~\eqref{eq:chi:t:q}). At moderate values of $\Delta \tau{\approx} 1$, %(i.e., in the ballistic regime), 
the dispersion develops a quadratic dependence on $q$ (see Fig.~\ref{fig:fig2}(c)). The results of Refs.~\cite{AndrianovPopov,Phan2023} are recovered in the limit $\Delta \tau {\rightarrow} \infty$ (see \cite{SM} for a detailed analysis of the crossover between the dirty and clean regimes). A similar analysis of the SH susceptibility in the complex momentum plane at fixed $\bar{\omega}$ also reveals a pole \cite{SM}. \color{black}

%%%%%%%%%%%%%%%%%%%%%%%%%%%%%%%%%%%
\begin{figure*}[t!]
\center{\includegraphics[width=1.0\textwidth]{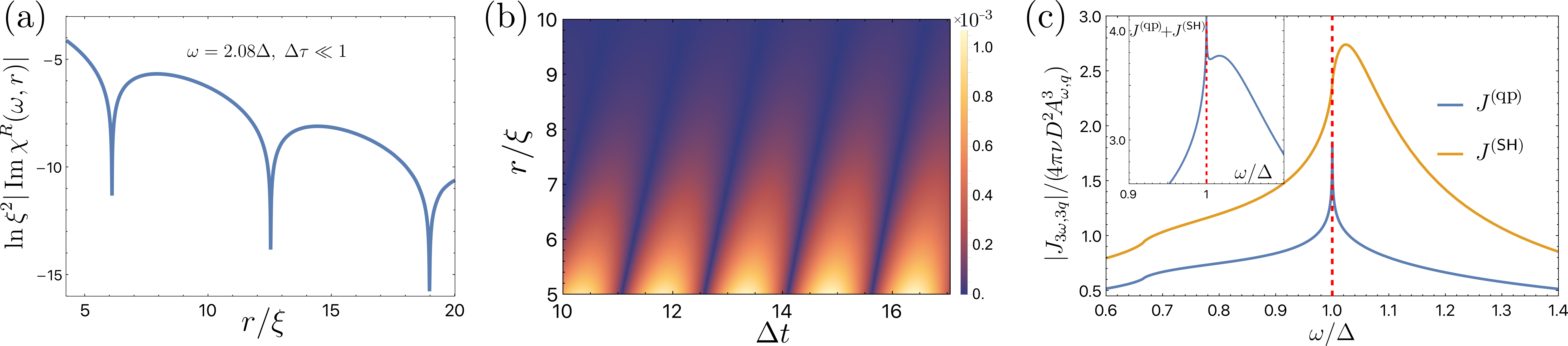}}
\caption{Oscillations in  (a) $\ln\xi^2|\operatorname{Im}\chi_{\textsf{SH}}^{R}(\omega,r)|$ and in (b) $|\xi^2\chi_{\textsf{SH}}^{R}(t,r)/\Delta|$ in 2D, and in the dirty limit $\Delta\tau{\ll} 1$. (c) The absolute value of the individual contributions to the current for
% $\xi q{=}0.5$ 
$Dq^2/\Delta{=}1/8$ for $\Delta\tau{\ll} 1$. The red dashed line indicates the $q{=}0$ resonance at $\omega{=}\Delta$. The blue (orange) curve corresponds to the quasiparticle (Schmid-Higgs) contribution. The inset shows the total current. Additional details on the dependence of the current on $Dq^2/\Delta$ are provided in Ref.~\cite{SM}. }  
\label{fig:fig3}
\end{figure*}
%%%%%%%%%%%%%%%%%%%%%%%%%%%%%%%%%%%

\noindent\textsf{\textcolor{blue}{Late-time and long-distance SH oscillations.}} 
Let us now discuss how the aforementioned pole manifests itself in various asymptotic limits of $\chi^R_{\textsf{SH}}$. First, we consider $\chi^R_{\textsf{SH}}(t,\bm{q}) $ at late times $t$ and fixed momentum, which describes a response to a sudden, spatially periodic perturbation. For arbitrary $\Delta\tau$, we find
\begin{align}
   \chi^R_{\textsf{SH}}(t,\bm{q}) &\simeq 2\operatorname{Im}[Z_{\bm{q}} e^{-i \omega_{\textsf{SH}}(\bm{q})t}]e^{-\gamma_{\textsf{SH}}(\bm{q})t} \theta(\omega_{\textsf{SH}}(\bm{q}){-}2\Delta)\notag \\
   &-\frac{2\sin (2\Delta t)}{\pi t^2} \partial_\omega \operatorname{Im}\chi^{R}_{\textsf{SH}}(\omega,\bm{q})_{\omega= 2\Delta+0^+}\;,
   \label{eq:chi:t:q}
\end{align} 
where  $Z_{\bm{q}}$ is the residue of $\chi^\downarrow_{\textsf{SH}}(z,\bm{q})$ at $z_{\bm{q}}$. The asymptotic late-time behavior is determined by the second term stemming from the continuum edge at $\omega{=}2\Delta$ in $\operatorname{Im}\chi^R_{\textsf{SH}}$. These oscillations at frequency $2\Delta$ 
decay as $1/t^2$, in contrast to the conventional ${\sim}1/\sqrt{t}$ decay at $q{=}0$ \cite{vaks1962collective}. The first term in Eq.~\eqref{eq:chi:t:q} originates from the pole $z_{\bm{q}}$, provided that its real part, $\omega_\textsf{SH}(\bm{q})$, exceeds $2\Delta$. Although this exponentially decaying contribution is sub-leading, interestingly, its frequency is $q$-dependent. In the dirty limit, the pole is ``hidden'' by the branch cut ($\omega_{\textsf{SH}}(\bm{q}){<}2\Delta$, see Fig.~\ref{fig:fig2}(a,b)) so its contribution to $\chi^R_{\textsf{SH}}$ is strongly incoherent and does not produce oscillations. Thus for $\Delta\tau{\ll}1$ and $\xi q{\ll}1$, Eq.~\eqref{eq:chi:t:q} yields 
\begin{equation}\label{eq:osc_t_q}
  \chi^R_{\textsf{SH}}(t,\bm{q}) {\approx} {-} \sin (2\Delta t) / [\Delta t^2 (\xi q)^6\ln^2(\xi q/2)] \;.
\end{equation}
At moderate disorder (or large momentum), the SH pole shifts into the right half-plane ($\omega_{\textsf{SH}}{>}2\Delta$) and becomes ``visible'' (cf. Fig.~\ref{fig:fig2}(c)), resulting in a coherent feature in $\chi^R_{\textsf{SH}}$ and additional oscillations (first term in Eq.~\eqref{eq:chi:t:q}). The ``critical'' $q_{c}$ at which these oscillations first appear (i.e. $\omega_{\textsf{SH}}(q_{c}){=}2\Delta$) is analyzed in \cite{SM}. 

The long-distance oscillations of $\chi^R_{\textsf{SH}}(\omega,\bm{r})$ at a fixed frequency $\omega$ are closely related to the pole in $\chi_{\textsf{SH}}^\downarrow$ in the complex momentum space. These oscillations correspond to a spatially-local periodic external drive and exist for $\omega{>}2\Delta$ only. In 2D and in the dirty limit, $\Delta\tau{\ll} 1$, we find 
\begin{equation}\label{eq:omega-r-FT}
 \operatorname{Im}\chi^R_{\textsf{SH}}(\omega,\bm{r}){\simeq} \frac{2^{1/4}\sqrt{\xi_\omega/r}}{\sqrt\pi\xi^{2} \ln \frac{ 2^4\Delta}{\omega{-}2\Delta}} e^{{-}{r}/{\xi_\omega}}\sin\left(\frac{r}{\xi_\omega}{+}\frac{\pi}{8}\right) ,
\end{equation}
where $\xi_\omega {=} \xi [|\ln((\frac{\omega}{\Delta}{-}2)/2^4)|^2/(\pi^2 (\frac{\omega}{\Delta}{-}2))]^{1/4}$ is the  period of oscillations diverging at the threshold $\omega{=}2\Delta$.

The long-distance and late-time behavior of $\chi_{\textsf{SH}}(t,\bm{r})$ in the regime $\Delta\tau{\ll} 1$ is sub-diffusive, featuring oscillations as a function of %$r^4/t \ln^2(\Delta t) $ 
{ $\varkappa{=}\pi^2 (r/\xi)^4/(\Delta t)\ln^2 (\Delta t)$}
with an approximate dynamical exponent $z{=}4$ (which implies the relation $r{\sim} t^{1/z}$). 
In 2D, for $\varkappa{\gg}1$, we find
\begin{equation}\label{eq:z=4_oscillations}
    \chi^R_{\textsf{SH}}(t,\bm{r}) \simeq \frac{2^{3/4} e^{{-}\frac{3\sqrt{3}}{8}\varkappa^{1/3}} }{\sqrt{3}\pi \xi^{2}  t \ln (\Delta t)}\cos\left(\frac{3}{8}\varkappa^{1/3}{-}2\Delta t\right)\; .
\end{equation}
%for large $\varkappa{=}\pi^2 (r/\xi)^4/(\Delta t)\ln^2 (\Delta t)$. 
In the opposite regime, $\Delta\tau {\gg} 1$, the amplitude SH fluctuations propagate diffusively, with $z{=}2$. %This form of 
Oscillations \eqref{eq:z=4_oscillations} can be induced by a quenched perturbation affecting a local gap magnitude. The results in 3D are qualitatively similar to Eqs.~\eqref{eq:omega-r-FT}-\eqref{eq:z=4_oscillations} and are presented in \cite{SM}. The analytic results~\eqref{eq:omega-r-FT}-\eqref{eq:z=4_oscillations} are benchmarked against direct numerical evaluation of $\chi_{\textsf{SH}}^R$ shown in Fig.~\ref{fig:fig3}.

\noindent\textsf{\textcolor{blue}{SH mode and the nonlinear current response.}} The results for $\chi_{\textsf{SH}}$ outlined above can be directly probed via the electromagnetic response to an external harmonic vector potential $\bm{A}(t,\bm{r}){=} \bm{A}_{\omega,\bm{q}}e^{i \bm{q}\cdot\bm{r}{-}i\omega t}$. As well-known, at cubic order in  $\bm{A}_{\omega,\bm{q}}$, induced corrections to the amplitude fluctuations at frequency $2\omega$ and momentum $2\bm{q}$ lead to the current oscillations at frequency $3\omega$ and momentum $3\bm{q}$ -- the effect known as third harmonic generation (THG)~\cite{SH_measured_2014,Naoto2015,Cea2016,Murotani2019,Silaev2019}. After evaluating the diagrams familiar from the $q{=}0$ case \cite{Silaev2019}, we find that the resulting paramagnetic contribution to the current (which is absent in the clean case \cite{Cea2016,Shimano2019}) consists of two terms, $\bm{J}_{3\omega,3\bm{q}}{=}\bm{J}^{(\rm qp)}_{3\omega,3\bm{q}}{+}\bm{J}^{(\textsf{SH})}_{3\omega,3\bm{q}}$ (see End Matter). The first term corresponds to a direct quasiparticle (qp) channel, and the second one involves the SH susceptibility:
\begin{equation}\label{eq:current}
\bm{J}^{(\textsf{SH})}_{3\omega,3\bm{q}}=
4\pi \nu D^2 \chi^R_{\textsf{SH}}(2\omega, 2q)\mathcal{B}_{\textsf{SH}}^R(\omega,q) |\bm{A}_{\omega,\bm{q}}|^2 \bm{A}_{\omega,\bm{q}}.
\end{equation}
The lengthy expression for $\mathcal{B}_{\textsf{SH}}^R(\omega,q)$ is given in \cite{SM}. Unlike the SH susceptibility $\chi^R_{\textsf{SH}}(2\omega, 2q)$, $\mathcal{B}_{\textsf{SH}}^R(\omega,q)$ does not exhibit any sharp non-analytic features in the 
limit $q{\to}0$ and $\omega{\to}\Delta$ and it could be replaced in Eq.~\eqref{eq:current} with its limiting value $\mathcal{B}_{\textsf{SH}}^R(\Delta,0){\approx} {-}1.55{-}1.27i$. Consequently, for $|\omega{-}\Delta|{\ll} \Delta$, the current $\bm{J}^{(\textsf{SH})}_{3\omega,3\bm{q}}$ is essentially governed by the SH susceptibility and exhibits a peak at a frequency $\omega_{\rm max}(\bm{q})/2{>}\Delta$ with the height that scales as $1/(\xi q)^2$, see Eq.~\eqref{eq:susceptibility_peak_width} and Fig.~\ref{fig:fig3}(c). In contrast, we find that the peak in the ``quasiparticle" contribution $\bm{J}^{(\rm qp)}_{3\omega,3\bm{q}}$ remains fixed at $\Delta$ even for $q{\neq}0$, as shown in Fig.~\ref{fig:fig2}(c). Intuitively, this is expected because the momentum-resolved  collective dynamics of the order parameter fluctuations is very distinct from individual quasiparticle excitations. Since $\bm{J}^{(\rm qp)}_{3\omega,3\bm{q}}$ at $\omega{=}\omega_{\rm max}(\bm{q})/2$ grows as $\ln[1/(\xi q)]$ only (see End Matter), the peak in SH contribution dominates the quasiparticle contribution at $\xi q{\ll} 1$. The pair-breaking effects (e.g. magnetic impurities) will broaden both peaks, thereby establishing a lower bound on $q$ for observation of the SH peak in the THG.
Thus, the emergence of an additional peak at a frequency above $\Delta$ in the finite-momentum current provides an unambiguous signature of the amplitude SH fluctuations and allows for a direct investigation of their dynamics summarized by Eqs.~\eqref{eq:osc_t_q}-\eqref{eq:z=4_oscillations}. 
%{\color{red}We also note that in the absence of pair-breaking effects (e.g., magnetic impurities), the peaks of the quasiparticle and SH contributions remain well separated in the limit $\xi q{\ll}1$. This is because the quasiparticle contribution, $\bm{J}^{(\rm qp)}_{3\omega,3\bm{q}}$, increases only as $\ln (1/\xi q)$ (see End Matter) at the position of the peak in $\bm{J}^{(\textsf{SH})}_{3\omega,3\bm{q}}$, while the SH contribution grows as $1/(\xi q)^2$. However, we expect that pair-breaking effects will broaden both peaks, thereby establishing a lower bound on the accessible momentum range. }

\color{black}

\noindent\textsf{\textcolor{blue}{Conclusions.}} In this work, we studied the spatially-resolved dynamics of the order parameter amplitude (SH) fluctuations in BCS superconductors with non-magnetic impurities. We identified a pole on the unphysical Riemann sheet of the SH susceptibility, associated with the oscillatory mode exhibiting sub-diffusive  $z{=}4$ spreading in the dirty limit. This pole also produces a peak in the spectral function above the edge of the two-particle continuum, even though the frequency of the SH mode itself can be below $2\Delta$ for sufficiently strong disorder. We also calculated the contribution of the SH mode to the nonlinear current response, focusing on the THG.
%the third harmonic generation (THG). 
We found that at finite wave vectors, the THG current exhibits an additional peak in its amplitude, shifted away from the conventional resonance at $\omega{=}\Delta$. This extra peak arises solely from the dynamics of the amplitude SH mode. Importantly, both disorder and finite wave vectors are essential for this effect. Without disorder, the SH mode would not contribute to the current \cite{Cea2016}, and at zero wave vector, multiple processes conflate into a single peak at $\Delta$ \cite{Silaev2019}, making it difficult to disentangle the SH contribution.
Our findings could be directly tested with spatially-resolved terahertz and
Raman spectroscopic probes \cite{SHTerahertz2013,SH_measured_2014,sherman2015higgs,Grasset2018}. Moreover, a finite wave vector can be imprinted in the current response in thin films with high-frequency surface acoustic waves \cite{Willett1993,Willett1993_2}, diffraction gratings and micropatterning \cite{Mackenzie2021}, finite spot size of the pump pulse \cite{Shimano2019,Dyke2024}, or by adding extra layers of a 2D van der Waals material with a Moiré superlattice. Another promising possibility is to use spatially-inhomogeneous Feshbach modulation of the interaction strength in disordered cold gases \cite{Carusotto2005} to directly quench the local value of the gap and study its spatial relaxation \cite{Endres2012}. 
%In the future, we plan to extend our analysis by including spatial fluctuations of the mean-field order parameter, $\delta\Delta(\bm{r})$, which can smear the coherence peaks and induce sub-gap states \cite{larkin1972density,Meyer2001,Feigelman2012,Abhisek2020}, leading to additional non-analyticities in the SH susceptibility below the continuum edge. 

	\textbf{Acknowledgments.\,--} We thank A.~Chubukov, A. Mel'nikov, V.~Kravtsov, S.~Raghu, and A. Levchenko for fruitful discussions.  The work of P.A.N. was supported in part by the US Department of Energy, Office of Basic Energy Sciences, Division of Materials Sciences and Engineering, under contract number DE-AC02-76SF00515. The work of I.S.B. was supported by the Russian Ministry of Science and Higher Education and by the Basic Research Program of HSE.
	I.S.B. acknowledges personal support from the Foundation for the Advancement of Theoretical Physics and Mathematics ``BASIS''.
 
		\bibliography{bibSH}

         \vspace{10em}
 \onecolumngrid
        \begin{center}
             {\large \textbf{End Matter}}
         \end{center}
 \twocolumngrid
\renewcommand{\theequation}{A\arabic{equation}}
\setcounter{equation}{0}
 \noindent\textsf{\textcolor{blue}{Appendix A: formula for $\chi^{\downarrow}_{\textsf{SH}}(z,\bm{q})$}}.
 The explicit formula for the function $\chi^{\downarrow}_{\textsf{SH}}(z,\bm{q})$ is given by
\begin{equation}\label{eq:Chi_SH_down}
\frac{1}{\chi^{\downarrow}_{\textsf{SH}}(z,\bm{q})} {=} \begin{cases}
[\chi_{\textsf{SH}}(z,\bm{q})]^{-1},\quad\operatorname{Im}z{>}0\\
[\chi_{\textsf{SH}}(z,\bm{q})]^{-1} {+}2 i \rho_{\bm{q}}(z),\quad \operatorname{Im}z{\leq} 0. 
    \end{cases}
\end{equation}
Here $[\chi_{\textsf{SH}}(z,\bm{q})]^{-1}$ is defined in Eq.~\eqref{eq:SH_complex}, and $\rho_{\bm{q}}(z) $ is the analytic continuation of $\rho_{\bm{q}}(\omega) $, given in Eq.~\eqref{eq:Pi_c_m_im} for $\Delta\tau{\ll} 1$ and for arbitrary $\Delta\tau$  in \cite{SM}, from $\omega{\geq} 2\Delta$ into the lower complex half-plane. 

\renewcommand{\theequation}{B\arabic{equation}}
\setcounter{equation}{0}
 \noindent\textsf{\textcolor{blue}{Appendix B: SH pole for $\Delta\tau{\ll}1$}}.
Next, we analytically derive the expression for the pole in the dirty limit. Instead of using our global integral representation in Eq.~\eqref{eq:Chi_SH_down}, we will follow an equivalent route by analytically continuing the approximate expression for $\chi^R_{\textsf{SH}}(\omega,\bm{q})$ (given in Eq.~\eqref{eq:chi_approx_cont} on the real axis at $\omega{>}2\Delta$) into the lower half plane, thereby ensuring the smoothness of the resulting function across the cut. After setting the inverse of the r.h.s. of Eq.~\eqref{eq:chi_approx_cont} to zero and treating $u$ as a complex variable, we obtain the following equation 
$\ln(u^* \bar{q}^2/8){=} i\pi (1{-}u^*)/2$. 
Its solution is given by 
$u^*{\simeq} 2{+}2i[1{-}W\left (4\pi /\bar{q}^2\right )]/\pi$, 
where 
% $c{\equiv} \sqrt{\pi} e^{c_1}{\approx}3.54$,
% and 
$W(y)$ is the Lambert function defined as a solution of the equation 
$W\exp W {=} y$. For $y{\gg} 1$, we find $W(y){\simeq} \ln(y/\ln y)$, and thus we can assume that $W(4\pi /\bar{q}^2) {\gg} 1$. The resulting behavior near the pole is given by
\begin{equation} \label{eq:pole}
\chi^\downarrow_{\textsf {SH}}(z,\bm{q}) \simeq \frac{Z_{\bm{q}}}{z-z_{\bm{q}}},\quad \frac{Z_{\bm{q}}}{\Delta}\simeq\frac{4\bar{q}^2}{\pi^2}\bigg[W\left(\frac{4\pi }{\bar{q}^2}\right)+i\pi\bigg],
\end{equation}
and the position of the pole $z_{\bm{q}}$ is given by
\begin{equation}
\label{eq:z_q_final}
    \frac{z_{\bm{q}}}{\Delta} \simeq 2 {-} \frac{  \bar{q}^4}{\pi^2}\left[W (4\pi/\bar{q}^2){-}1\right]^2 {-} \frac{2i \bar{q}^4}{\pi} \left[W(4\pi/\bar{q}^2){-}1\right].
\end{equation}
Within the leading logarithmic accuracy, we obtain Eq.~\eqref{eq:dispersion_damping}.

The position of the pole in the complex momentum plane can also be found from Eq.~\eqref{eq:chi_approx_cont}. After rewriting the expression in Eq.~\eqref{eq:chi_approx_cont} as
\begin{equation}
\chi_{\textsf{SH}}^R(\omega,\bm{q}) {\simeq} 2\sqrt{\frac{u^2{-}1}{ \bar{\omega}{-}2}}\bigr[  \ln \frac{2^4}{\bar{\omega}{-}2}
{+}u \ln\frac{u{-}1}{u{+}1}{+}i\pi (1{-}u)\bigr]^{-1} \end{equation}
for $\omega{>}2\Delta$ and $\Delta\tau{\ll}1$, 
setting the inverse of this expression to zero, finding the solution $\tilde{u}_{*}$, and converting it back to momentum $\bar{q}$, we obtain

\begin{equation}\label{eq:SH_pole_momentum}
\chi_{\textsf{SH}}^R(\omega,\bm{q}){\simeq} \frac{\tilde{Z}_\omega}{\bar{q}^2{-}\bar{q}_\omega^2},\quad
    \bar{q}^2_\omega {\simeq} \frac{2\pi  \sqrt{\bar{\omega}{-}2}}{ \ln \frac{2^4}{\bar{\omega}{-}2}}\bigg[i {+}\frac{\pi }{\ln \frac{2^4}{\bar{\omega}{-}2}}\bigg],
\end{equation}
and $\tilde{Z}_\omega {\simeq} 1/|\ln((\bar{\omega}{-}2)^{1/4}/2)|$
is the residue at $\bar{q}^2_\omega$. Its Fourier transform leads to Eq.~\eqref{eq:omega-r-FT} and \eqref{eq:z=4_oscillations} \cite{SM}.

\renewcommand{\theequation}{C\arabic{equation}}
\setcounter{equation}{0}
 \noindent\textsf{\textcolor{blue}{Appendix C: explicit expressions for the nonlinear current response.}} The nonlinear current response at  $q{=}0$ was analyzed in \cite{Silaev2019}. Here, we extend that analysis to finite $q$. The diagrams determining the paramagnetic contribution to the current are identical to those in \cite{Silaev2019}.
% and are reproduced in Fig.~\ref{fig:diagrams}. 
A straightforward calculation yields Eq.\eqref{eq:current}, where the function $\mathcal{B}_{\textsf{SH}}^R(\omega,q)$ is expressed as a product of two fermionic loops, each connecting two external vector potential vertices to the SH susceptibility
\begin{equation}
\mathcal{B}_{\textsf{SH}}^R(\omega,q) = -2\pi B_{1}^R(\omega,2q)B_{2}^R(\omega,2q)\;,
\end{equation}
The expressions for $B_{1}(\omega_n,q)$ and $B_{2}(\omega_n,q)$ on the Matsubara axis are given in \cite{SM}. Here, we only provide the result after analytic continuation for $\Delta\tau{\ll}1$ and at $T=0$
\begin{equation}
\begin{aligned}\label{eq:B1}
    \frac{B_{1}^R(\omega,q)}{\Delta }&=\int\limits_{0}^\omega \frac{d\varepsilon}{2\pi i} F_{\omega,q}^{RA}(\varepsilon)+\int\limits_{-\infty}^{+\infty} \frac{d\varepsilon}{4\pi i} \operatorname{sgn}(\varepsilon{-}\omega)F_{\omega,q}^{RR}(\varepsilon),\\
    F_{\omega,q}^{Rs}(\varepsilon)&=\frac{\Delta^2+ 3\varepsilon^2-\omega^2-E^R_{\varepsilon+\omega}E^s_{\varepsilon-\omega}}{(Dq^2+E^R_{\varepsilon+\omega}+E^s_{\varepsilon-\omega})E^R_\varepsilon E^R_{\varepsilon+\omega}E^s_{\varepsilon-\omega}},\\
    E^{R}_\varepsilon=& \theta(\Delta{-}|\varepsilon|)\sqrt{\Delta^2{-}\varepsilon^2}- i\theta(|\varepsilon|{-}\Delta) \operatorname{sgn}(\varepsilon)\sqrt{\varepsilon^2{-}\Delta^2}
    \end{aligned}
\end{equation}
and $E^A_\varepsilon = E^R_{-\varepsilon}$. Similarly, for $B_2^R$, we find

\begin{equation}
\begin{aligned}\label{eq:B_2}
    \frac{B_{2}^R(\omega,q)}{\Delta }&=\int\limits_{0}^\omega \frac{d\varepsilon}{2\pi i} \Phi_{\omega,q}^{RAA}(\varepsilon)+ \int\limits_{\omega}^{2\omega} \frac{d\varepsilon}{2\pi i} \Phi_{\omega,q}^{RRA}(\varepsilon)\\
    &+ \int\limits_{-\infty}^{+\infty} \frac{d\varepsilon}{4\pi i} \operatorname{sgn}(\varepsilon{-}2\omega)\Phi_{\omega,q}^{RRR}(\varepsilon)\;,
  \end{aligned}
\end{equation}
where we also defined
\begin{equation}
\begin{aligned}\label{eq:B2}
    \Phi_{\omega,q}^{Rs \sigma}(\varepsilon)&=\frac{1}{(Dq^2+E^R_{\varepsilon+\omega}+E^s_{\varepsilon-\omega})E^R_{\varepsilon+\omega}E^s_{\varepsilon-\omega}}\\
    &\times \bigg\{ \bigg(\frac{1}{E^R_{\varepsilon+2\omega}}{+}\frac{1}{E^\sigma_{\varepsilon-2\omega}}\bigg)  \big(\Delta^2{+}\varepsilon^2{-}\omega^2{-}E^R_{\varepsilon+\omega}E^s_{\varepsilon-\omega}\big)\\
    &+2\varepsilon\bigg(\frac{\varepsilon+2\omega}{E^R_{\varepsilon+2\omega}}+\frac{\varepsilon-2\omega}{E^\sigma_{\varepsilon-2\omega}}\bigg)\bigg\}\;.
    \end{aligned}
\end{equation}
Finally, the quasiparticle contribution, $J^{(\rm qp)}_{3\omega,3q}$, is given by 
 % the diagrams in Fig.\ref{fig:diagrams}(c). We obtain the following expression
\begin{equation}
    J^{(\rm qp)}_{3\omega,3q}= -4\pi \nu D^2 B_3^R(\omega,2q)|\bm{A}_{\omega,\bm{q}}|^2 \bm{A}_{\omega,\bm{q}}, 
    \end{equation}
    where $B_3(\omega_n, q)$ on the Matsubara axis is provided in \cite{SM}. The real frequency expression for $B_3^R(\omega,q)$ is the same as for $B_2^R(\omega,q)$ in Eq.~\eqref{eq:B_2}, but with $\Phi_{\omega,q}^{Rs \sigma}(\varepsilon)$ replaced by $   W_{\omega,q}^{Rs \sigma}(\varepsilon)$, where  
    \begin{equation}
\begin{aligned}\label{eq:B3}
    W_{\omega,q}^{Rs \sigma}(\varepsilon)&=\frac{1}{(Dq^2{+}E^R_{\varepsilon+\omega}{+}E^s_{\varepsilon-\omega})E^R_{\varepsilon}}\bigg\{ \bigg(\frac{\Delta}{E^R_{\varepsilon+2\omega}}{+}\frac{\Delta}{E^\sigma_{\varepsilon-2\omega}}\bigg) \\
    & \times \bigg(1+\frac{\omega^2{-}\Delta^2{-}3\varepsilon^2}{E^R_{\varepsilon+\omega}E^s_{\varepsilon-\omega}}\bigg)   -
\frac{\varepsilon}{\Delta}  \bigg(1+\frac{3\Delta^2{-}\omega^2{+}\varepsilon^2}{E^R_{\varepsilon+\omega}E^s_{\varepsilon-\omega}}\bigg)\\
    & \times
    \bigg(\frac{\varepsilon+2\omega}{E^R_{\varepsilon+2\omega}}+\frac{\varepsilon-2\omega}{E^\sigma_{\varepsilon-2\omega}}\bigg)\bigg\}.
    \end{aligned}
\end{equation}
Near the resonance at $\omega{\approx}\Delta$, we find a logarithmic divergence
\begin{equation}
    B_3^R(\omega,q)\approx \frac{1}{\pi}\frac{1}{Dq^2/\Delta +1-i\sqrt{3}}\ln \frac{\Delta}{|\omega-\Delta|}\;.
\end{equation}
Therefore, $ J^{(\rm qp)}_{3\omega_{\rm max}(q)/2,3q}$ increases as $\ln1/(\xi q)$ for $\xi q{\ll}1$. The expressions for arbitrary $\Delta\tau$ can be obtained by replacing factors $(Dq^2+E^R_{\varepsilon+\omega}+E^s_{\varepsilon-\omega})^{-1}$ in Eqs.~(\ref{eq:B1}), (\ref{eq:B2}), (\ref{eq:B3}) with $S_{\bm{q}}\left(E^R_{\varepsilon+\omega}+E^s_{\varepsilon-\omega}\right)$.

\foreach \x in {1,...,10} 
{% 
\clearpage 
\includepdf[pages={\x},turn=false]{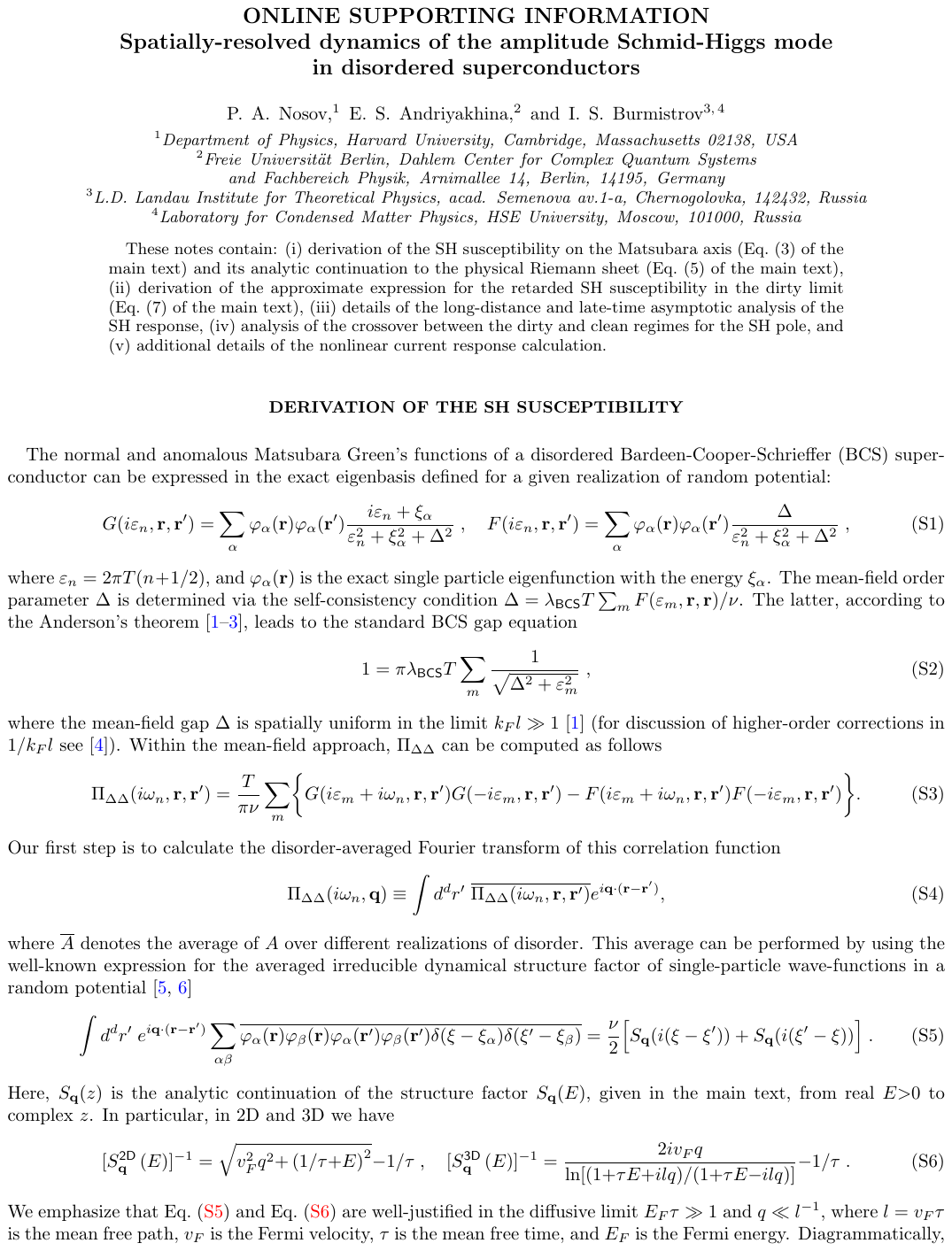}
}

  \end{document}